\documentclass[showpacs,preprintnumbers,superscriptaddress]{revtex4}
\usepackage{CJK}
\usepackage{amsmath,amssymb,graphicx,bm}
\begin{document}

\title{Thermodynamics of black hole in $D$-dimensional $f(R)$ theory}

\author{Chenrui Zhu}
\affiliation{College of Physical Science and Technology, Hebei University, Baoding 071002, China}
\author{Rong-Jia Yang \footnote{Corresponding author}}
\email{yangrongjia@tsinghua.org.cn}
\affiliation{College of Physical Science and Technology, Hebei University, Baoding 071002, China}
\affiliation{Hebei Key Lab of Optic-Electronic Information and Materials, Hebei University, Baoding 071002, China}
\affiliation{National-Local Joint Engineering Laboratory of New Energy Photoelectric Devices, Hebei University, Baoding 071002, China}
\affiliation{Key Laboratory of High-pricision Computation and Application of Quantum Field Theory of Hebei Province, Hebei University, Baoding 071002, China}

\begin{abstract}
We consider whether the new horizon-first law works in higher-dimensional $f(R)$ theory. We firstly obtain the general formulas to calculate the entropy and the energy of a general spherically-symmetric black hole in $D$-dimensional $f(R)$ theory. For applications, we compute the entropies and the energies of some black hokes in some interesting higher-dimensional $f(R)$ theories.
\end{abstract}
\maketitle
\section{Introduction}
Since Bekenstein's and Hawking's work \cite{Bekenstein:1973ur, Hawking:1974sw}, it is convinced that there may be a deep relation between the gravitational field equations and the laws of thermodynamics. Like in thermodynamics, four laws of black hole dynamics were found in \cite{Bardeen:1973gs}. The field equations of general relativity in its tensorial form can be derived by applying the Clausius relation $\delta Q = T\delta S$ on the horizon of spacetime, here $\delta Q$ is the energy flux across the horizon and $\delta S$ and $T$ are the change in the entropy and the Unruh temperature seen by an accelerating observer just inside the horizon \cite{Jacobson:1995ab}. From Einstein equations, one can obtain the thermal entropy density of spacetime without assuming the temperature or the horizon \cite{Yang:2011sx, Yang:2014kna}. It was shown that for a generalized gravity theory, the field equations are equivalent to the first law of thermodynamics \cite{Brustein:2009hy}. This programme was also applied to other modified gravity theories: such as $f(R)$ theory \cite{Eling:2006aw, Elizalde:2008pv}, and scalar-Gauss--Bonnet gravity \cite{Bamba:2009gq}. It was shown, however, that the Bekenstein--Hawking entropy depends not only on the black hole parameter, but also on the coupling which induces Lorentz violation \cite{Chen:2015hsm}.

For a spherically-symmetric spacetime, Einstein's field equations can be written in the form of thermodynamic identity (called the horizon-first law): $dE=TdS-PdV$ \cite{Padmanabhan:2002sha}. This framework of horizon thermodynamics has also been extended to other theories of gravity \cite{Paranjape:2006ca, Sheykhi:2014rka} and the non-spherically-symmetric cases \cite{Kothawala:2007em}. The horizon-first law, however, has two shortcomings: (a) the thermodynamic variables are vague in the original derivation and require further determination, (b) both $S$ and $V$ are functions of only $r_+$, so does the horizon-first law, which makes the terms `heat' and `work' confused \cite{Hansen:2016gud}. To avoid these two problems, a new horizon-first law was proposed in \cite{Hansen:2016gud}, where the temperature $T$ and the pressure $P$ are independent thermodynamic quantities and the entropy and free energy are derived concepts, while the horizon-first law can be restored by the Legendre projection. This procedure was generalized to in $f(R,R^{\mu\nu}R_{\mu\nu})$ theory \cite{Feng:2019ejb} and $f(R)$ theory with a spherically-symmetric black hole \cite{Zheng:2018fyn} or with a general spherically-symmetric black hole \cite{Zheng:2019mvn}.

The natural generalization of general relativity is the higher-dimensional and higher-order gravity. Higher-dimensional black hole in higher-dimensional gravity is physically interesting, whose physics are markedly different and much richer than those in four dimensions, see, for example, there are limits on the ratio of mass to charge for Tangherlini--Reissner--Nordstrom black hole \cite{Yang:2018cim}. Extra dimensions are also needed for consistency in string theory. So it is valuable to study the physics of high-dimensional black holes. In addition, energy issue in higher-dimensional and higher-order gravity is still an open problem, some attempts to find a satisfactory answer to this problem have been proposed \cite{Deser:2002jk, Deser:2007vs, Cai:2009qf, Abreu:2010sc, Cognola:2011nj}. In literature, there have been several attempts to define the concept of energy using local or quasi-local concepts, however, not all these definitions of energy agree with each other. Here we will investigate this issue in $D$-dimensional $f(R)$ black hole and hope to give interesting suggestions.

As one of the simplest modifications to general relativity, $f(R)$ gravity have been extensively studied over the past decade \cite{Sotiriou:2008rp,DeFelice:2010aj,Capozziello:2011et}. It's Lagrangian is a function of Ricci scalar in which higher-order terms can encapsulate high-energy modifications to general relativity, but the field equations are simple enough that it is possible to solve them. Secondly, and most importantly, $f(R)$ gravity does not suffer from Ostr\"{o}gradsky instability. Various applications of $f(R)$ gravity to cosmology have been investigated, such as inflation, dark energy, cosmological perturbations, and black hole solutions. Here we will investigate whether the new horizon-first law still works in higher-dimensional $f(R)$ gravity. We will adopt the method presented in \cite{Padmanabhan:2002sha} to define the energy and the entropy: writing the radial component of gravitational field equations on the horizon as the equation of state $P=D(r_+)+C(r_+)T$, which can be rewritten as a thermodynamic identity $\delta G=-S\delta T+V\delta P$, then identifying $S$ as the entropy, and taking $E=G+TS-PV$ as the energy. We will show that the new horizon-first law can give not only the entropy but also the energy of black hole in higher-dimensional $f(R)$ theory, which for some special cases are consistent with the results obtained by using other methods.

The structure of this paper is as follows. In Section 2, we briefly review the new horizon-first law and its applications in $f(R)$ theories. In Section 3, we consider whether the new horizon-first law still holds in higher-dimensional $f(R)$ theory. In Section 4, we discuss applications for some $D$-dimensional $f(R)$ theories. Conclusions and discussions are given in Section 5.

\section{The new horizon first law and its application in $f(R)$ theory}
Inspired by the radial Einstein equation on the horizon of Schwarzschild black hole, it is reasonable to suggest that the radial field equation of a gravitational theory under consideration takes the following form \cite{Hansen:2016gud}
\begin{eqnarray}
\label{p}
P=D(r_+)+C(r_+)T,
\end{eqnarray}
where $C$ and $D$ are functions of the radius of black hole, $r_+$, in general they depend on the gravitational theory one considered. The temperature $T$ in (\ref{p}) is identified from thermal quantum field theory, which is independent of any gravitational field equations \cite{Hansen:2016gud}. According to the conjecture proposed in \cite{Yang:2014kna}, the pressure in (\ref{p}) is identified as the $(^r_{r})$ component of the matter stress-energy, it also does not fall back on any gravitational field equations. Considering a virtual displacements $\delta r_+$ and varying the equation (\ref{p}), then multiplying the volume of black hole $V(r_+)$, yields \cite{Hansen:2016gud}
\begin{eqnarray}
\label{nhfl}
\delta G=-S\delta T+V\delta P,
\end{eqnarray}
comparing with the thermodynamical identity $\delta G=-S\delta T+V\delta P$, where $G$ can be identified as the Gibbs free energy which is given by \cite{Hansen:2016gud}
\begin{eqnarray}
\label{10}
G&=&\int V(r_+)D'(r_+)\,dr_{+}+T\int V(r_+)C'(r_+)dr_+ \nonumber\\
&=&PV-ST-\int V'(r_+)D(r_+)dr_+,
\end{eqnarray}
and $S$ is identified as the entropy which is \cite{Hansen:2016gud}
\begin{eqnarray}
\label{11}
S=\int V'(r_+)C(r_+)dr_+.
\end{eqnarray}

Using the degenerate Legendre transformation as that in thermodynamics, the energy $E$ is defined as $E=G+TS-PV$ which can be easily got \cite{Zheng:2018fyn}
\begin{eqnarray}
\label{energy}
E=-\int V'(r_+)D(r_+)dr_+.
\end{eqnarray}

This procedure was firstly investigated for Einstein gravity and Lovelock gravity which only give rise to second-order field equation \cite{Hansen:2016gud} and was also applied to $f(R)$ gravity with a general static spherically-symmetric black hole in $f(R)$ gravity
\begin{eqnarray}
\label{metric2}
ds^2=-W(r)dt^2+\frac{dr^2}{N(r)}+r^2 d\Omega^2,
\end{eqnarray}
where $W(r)$ and $N(r)$ are general functions of the coordinate $r$ and the event horizon is local at the largest positive root of $N(r_+)=0$ with $N'(r_+)\neq 0$, the entropy in this case is given by \cite{Zheng:2019mvn}
\begin{eqnarray}
\label{entropy2}
S&=&\int (2\pi r_+F+\pi r^2_+ F')dr_+=\pi r^{2}_{+}F,
\end{eqnarray}
where $F=\frac{df}{dR}$. The energy is found to be \cite{Zheng:2019mvn}
\begin{eqnarray}
\label{energy2}
E=\frac{1}{2}\int \sqrt{\frac{W'}{N'}}\left[\frac{F}{r^2_+}+\frac{1}{2}(f-RF)\right]r^2_{+}dr_+.
\end{eqnarray}
For $W(r)=N(r)$, Eq. (\ref{energy2}) reduces to the result obtained in \cite{Zheng:2018fyn} which is consistent with the expression obtained in \cite{Cognola:2011nj} and can be derived by using the unified first law of black hole dynamics \cite{Hayward:1997jp}; Equation (\ref{entropy2})is consistent with the results derived by using the Wald entropy formula or the Euclidean semiclassical approach \cite{Vollick:2007fh,Dyer:2008hb, Iyer:1995kg}. In the next section, we will consider the new horizon-first law in $D$-dimensional $f(R)$ Theory with a general static spherically-symmetric black hole.

\section{The entropy and energy of black hole in $D$-dimensional $f(R)$ theory}
In this section, we turn our attention to discussing whether the new horizon-first law still holds in the $D$-dimensional $f(R)$ theory. Considering a general spherically-symmetric and static $D$-dimensional black hole in $f(R)$ theory, its geometry is given by
\begin{eqnarray}
\label{metric3}
ds^2=-W(r)dt^2+\frac{dr^2}{N(r)}+r^2 d\Omega^2_{D-2},
\end{eqnarray}
in which $d\Omega^2_{D-2}$ represents the $D-2$-dimensional unit spherical line element. For the metric (\ref{metric3}), the surface gravity takes the form \cite{DiCriscienzo:2009hd}:
$\kappa_K=\sqrt{W'(r_+) N'(r_+)}/2$, giving the temperature of the black hole as
\begin{eqnarray}
\label{temp}
T=\frac{\kappa_K}{2\pi}=\frac{\sqrt{W'(r_+)N'(r_+)}}{4\pi}.
\end{eqnarray}

The action of $D$-dimensional $f(R)$ gravity with source is represented by
\begin{eqnarray}
\label{17}
I=\int \text{d}^{D}x\sqrt{-g}\left[\frac{f(R)}{2k^2}+L_{\rm m}\right],
\end{eqnarray}
where $k^2=8\pi$ and $D\geq 3$. Here we take the units $G=c =\hbar= 1$. $f(R)$ is a function of the Ricci scalar $R$ and $L_{\rm m}$ is the matter Lagrangian. Physically $f(R)$ theory must fulfil two stability conditions \cite{Pogosian:2007sw}: (a) no ghosts, $df/dR>0$; and (b) no tachyons, $d^2f/dR^2>0$ \cite{Dolgov:2003px}. Variation of the action (\ref{17}) with respect to metric provides the gravitational field equations
\begin{eqnarray}
\label{18}
G^{\nu}_{\mu}\equiv R^{\nu}_{\mu}-\frac{1}{2}\delta^{\nu}_{\mu}R=k^2\left(\frac{1}{F}T^{\nu}_{\mu}+\frac{1}{k^2}\mathcal{T}^{\nu}_{\mu}\right),
\end{eqnarray}
where $T_{\mu\nu}=\frac{-2}{\sqrt{-g}}\frac{\delta L_{\rm m}}{\delta g^{\mu\nu}}$ the energy-momentum tensor of the matter. We define the stress-energy tensor of the effective curvature fluid as $\mathcal{T}^{\nu}_{\mu}$ which is given by
\begin{eqnarray}
\label{19}
\mathcal{T}^{\nu}_{\mu}=\frac{1}{F(R)}\left[\frac{1}{2}\delta ^{\nu}_{\mu}(f-RF)+\nabla_{\mu}\nabla^{\nu}F-\delta ^{\nu}_{\mu}\Box F\right],
\end{eqnarray}
where $\Box =\nabla ^{\lambda}\nabla _{\lambda}$. Assuming the metric (\ref{metric3}), we derive after some calculations by using of the relations $\Box F=\frac{1}{\sqrt{-g}}\partial_{\mu}[\sqrt{-g}g^{\mu\nu}\partial_{\nu}F]$ the $(^1_1)$ components of the Einstein tensor and the effective curvature fluid respectively
\begin{eqnarray}
\label{21}
G^{1}_{1}=\frac{1}{r^2}\left[\frac{(D-2)(D-3)(N-1)}{2}+\frac{(D-2)NW'r}{2W}\right],
\end{eqnarray}
and
\begin{eqnarray}
\label{20}
\mathcal{T}^{1}_{1}=\frac{1}{F(R)}\left[\frac{1}{2}(f-RF)-\frac{N}{2W}W'F'-\frac{D-2}{r}NF'\right],
\end{eqnarray}
where the prime stands for the derivative with respected to $r$. Taking the trace of Equation (\ref{18}), yields the relation
\begin{eqnarray}
\label{22}
RF(R)-\frac{D}{2}f(R)+(D-1)\Box F =k^2 T^{\nu}_{\nu},
\end{eqnarray}
where $T^{\nu}_{\nu}$ is the trace of the energy-momentum tensor. Substituting Equations (\ref{21}), (\ref{20}), and $T^r_r=P$ into Equation (\ref{18}), yields
\begin{eqnarray}
\label{23}
k^2 P=\frac{F(D-2)(D-3)}{2r^2}(N-1)-\frac{1}{2}(f-RF)+\frac{NF'(D-2)}{r}+\frac{(D-2)NW'F}{2Wr}+\frac{NW'F'}{2W}.
\end{eqnarray}

Thinking of $N(r_+)=0$ and the temperature (\ref{temp}) at the horizon, Equation (\ref{23}) reduces to
\begin{eqnarray}
\label{24}
P=-\frac{1}{8\pi}\left[\frac{F(D-3)(D-2)}{2r^2_+}+\frac{1}{2}\left(f-RF\right)\right]+\frac{1}{4}\sqrt{\frac{N'}{W'}}\left[\frac{F(D-2)}{r_+}+F'\right]T.
\end{eqnarray}

Comparing Equations (\ref{24}) and (\ref{p}), we then get
\begin{eqnarray}
\label{25}
D(r_+)=-\frac{1}{8\pi}\left[\frac{F(D-3)(D-2)}{2r^2_+}+\frac{1}{2}\left(f-RF\right)\right],
\end{eqnarray}
and
\begin{eqnarray}
\label{26}
C(r_+)=\frac{1}{4}\sqrt{\frac{N'}{W'}}\left[\frac{F(D-2)}{r_+}+F'\right].
\end{eqnarray}

The volume $V$ of the black hole in $D$-dimensional spacetime takes the form \cite{Parikh:2005qs}
\begin{eqnarray}
\label{V}
V(r_+)=\frac{2\pi^\frac{D-1}{2}}{\Gamma(\frac{D-1}{2})}\int_{0}^{r_{+}}\sqrt{\frac{W(r)}{N(r)}}r^{D-2}dr.
\end{eqnarray}

Use the relation $\frac{N(r_+)}{W(r_+)}=\frac{N'(r_+)}{W'(r_+)}$ \cite{DiCriscienzo:2009hd}, we get
\begin{eqnarray}
\label{V1}
V'(r_{+})=\frac{2\pi^\frac{D-1}{2}}{\Gamma(\frac{D-1}{2})}\sqrt{\frac{W'(r_{+})}{N'(r_{+})}}r_{+}^{D-2}.
\end{eqnarray}

Substituting Equations (\ref{V1}) and (\ref{26}) into the expression (\ref{11}), the entropy of black hole (\ref{metric3}) in $D$-dimensional $f(R)$ gravity is
\begin{eqnarray}
\label{sdf}
S=\frac{\pi^\frac{D-1}{2}}{2\Gamma(\frac{D-1}{2})}\int [F(D-2)r_{+}^{D-3}+r^{D-2}_+ F']dr_+=\frac{\pi^\frac{D-1}{2}}{2\Gamma(\frac{D-1}{2})}Fr_{+}^{D-2}.
\end{eqnarray}

Inserting Equations (\ref{V1}) and (\ref{25}) into Equation (\ref{energy}), then we obtain the energy of black hole (\ref{metric3}) in $D$-dimensional $f(R)$ theory as
\begin{eqnarray}
\label{edf}
E=\frac{\pi^\frac{D-3}{2}}{4\Gamma(\frac{D-1}{2})}\int \sqrt{\frac{W'}{N'}}\left[\frac{F(D-2)(D-3)}{2r^2_+}+\frac{1}{2}(f-RF)\right]r^{D-2}_{+}dr_+.
\end{eqnarray}

Equations (\ref{sdf}) and (\ref{edf}) are the main results obtained in this work, they can be used to calculate the entropy and the energy of a specific black hole in a specific $D$-dimensional $f(R)$ gravity. For $D=4$, Equations (\ref{sdf}) and (\ref{edf}) recover Equations (\ref{entropy2}) and (\ref{energy2}). Fixing $f(R)=R$ and $W=N$, one expects the results to go back to the framework of higher-dimensional Einstein's gravity, see Equations (\ref{ssd}) and (\ref{esd2}) in the next section.

\section{Applications}
In this section, we will illustrate the procedure to calculate the entropy and the energy for black holes in a certain $f(R)$ theory by using Equations (\ref{sdf}) and (\ref{edf}). These models have solutions with constant Ricci curvature (such as a Schwarzschild or a Schwarzschild-de Sitter solution) or solutions with non-constant Ricci curvature.

\subsection{The constant Ricci curvature case}
We start with the simplest but important case, $F=1$, which implies $f=R-2\Lambda$ where $-2\Lambda$ is an integration constant to be  regarded as the cosmological constant. This model has a Schwarzschild or a Schwarzschild-de/anti de Sitter black hole solution \cite{Amirabi:2015aya,Bueno:2017sui}
\begin{eqnarray}
\label{sd}
W(r)=N(r)=\left\{
\begin{aligned}
&-M-\Lambda r^2,&~~~~~D=3,\\
&1-\frac{2M}{(D-3)r^{D-3}}-\frac{2\Lambda}{(D-2)(D-1)}r^{2},&~~~~D>3,\\
\end{aligned}
\right.
\end{eqnarray}
where $M$ is the mass of black hole. The solution is Schwarzschild-de/anti de Sitter black hole solution  for $D>3$ and it is the non-rotating BTZ black hole for $D=3$. The constant curvature $R_{0}$ from Equation (\ref{sd})
is given by
\begin{eqnarray}
\label{R0}
R_{0}=\frac{2D\Lambda}{D-2}.
\end{eqnarray}
From Equation (\ref{sdf}), the entropy is found to be
\begin{eqnarray}
\label{ssd}
S=\frac{\pi^{\frac{D-1}{2}}}{2\Gamma(\frac{D-1}{2})}r_{+}^{D-2}.
\end{eqnarray}

It reduces to $S=\pi r_+/2$ for a non-rotating BTZ black hole and returns to the standard results for $D=4$. The energy of the black hole is obtained from Equation (\ref{edf}) as
\begin{eqnarray}
 \label{esd1}
E=-\frac{1}{8}\Lambda r_{+}^{2}=\frac{1}{8}M,
\end{eqnarray}
where $-\Lambda r_{+}^{2}=M$ for $D=3$ has been used at the horizon. For $M=0$, it reduces to results presented in \cite{Padmanabhan:2002sha}. For $D>3$, it reads
\begin{eqnarray}
 \label{esd2}
E=\frac{\pi^\frac{D-3}{2}}{4\Gamma(\frac{D-1}{2})}\left[\frac{(D-2)r_{+}^{D-3}}{2}-\frac{\Lambda r_{+}^{D-1}}{D-1}\right]=\frac{\pi^{\frac{D-3}{2}}}{4\Gamma(\frac{D-1}{2})}\frac{D-2}{D-3}M,
\end{eqnarray}
where we used $N(r_+)=0$ for $D>3$ at the horizon. For 4-Dimensional Einstein's gravity, Equations (\ref{ssd}) and (\ref{esd2}) give $S=\pi r_+^2=A/4$ and
$E=M$, respectively. The nonnegativity of the energy, gives new constrains on the parameter: $\frac{(D-1)(D-2)}{2}>\Lambda r^2_+$ for $D>3$.

\subsection{The non-constant Ricci curvature case}
We apply the same procedure for black hole solutions with non-constant curvature which are more interesting. We consider two types of $f(R)$ theories: (a) $F$ is a linear function of $r$, and (b) $F$ is a power law function of $r$.

\subsubsection{$F(r)=1+\alpha r$}
In this case, $F$ is a linear function of $r$ with $\alpha$ a non-zero constant. $W(r)$ and $N(r)$ in three-dimensional spacetime are given by \cite{Amirabi:2015aya}
\begin{eqnarray}
W(r)=N(r)=C_2 r^2+C_1\left(\alpha r-\frac{1}{2}\right)-C_1 \alpha^2 r^2 \ln\left(1+\frac{1}{\alpha r}\right).
\end{eqnarray}

Function $f(R(r))$ reads
\begin{eqnarray}
f=-4C_2+4C_1\alpha^2 \ln\left(1+\frac{1}{\alpha r}\right)-\frac{2C_1\alpha(1+2\alpha r)}{r(1+\alpha r)},
\end{eqnarray}
where $C_i$ are integration constants with $C_1$ related to the mass of the central object and $C_2$ identified as the cosmological constant. The Ricci scalar $R$ evolves as
\begin{eqnarray}
R=-6C_2+6C_1\alpha^2 \ln\left(1+\frac{1}{\alpha r}\right)-\frac{C_1\alpha(2+9\alpha r+6\alpha^2 r^2)}{r(1+\alpha r)^2}.
\end{eqnarray}

Then, the entropy formula (\ref{sdf}) gives
\begin{eqnarray}
\label{S3}
S=\frac{\pi r_{+}}{2}(1+\alpha r_+),
\end{eqnarray}
gives limit on parameter: $\alpha\geq-1/r_+$ from $S\geq0$. The energy of the black hole is obtained from Equation (\ref{edf}) as
\begin{eqnarray}
\label{E3}
E&=&\frac{r_+^2}{8}\left[C_2+2\alpha^2 C_1+2\alpha C_2r_+-C_1\alpha^2(1+2\alpha r_+)\ln\left(1+\frac{1}{\alpha r_+}\right)\right]\\\nonumber
&=&\frac{C_1}{16},
\end{eqnarray}
where $N(r_+)=0$ was used. $E\geq 0$ gives a new constraint on the parameter: $C_1\geq 0$.

For $D=4$ spacetime, $W(r)$ and $N(r)$ take the forms
\begin{eqnarray}
 \label{WN3}
 W(r)=N(r)=C_2 r^2+\frac{1}{2}+\frac{1}{3\alpha r}+\frac{C_1}{r}\left[3\alpha r-2-6\alpha ^2r^2+6\alpha^3 r^3\ln\left(1+\frac{1}{\alpha r}\right)\right],
\end{eqnarray}
where $C_2$ is related to the cosmological constant. We note that Equation (\ref{WN3}) is different from Equation (27) in \cite{Amirabi:2015aya}. Function $f(R(r))$ is given by
\begin{eqnarray}
\label{fBD4}
f=-6C_2-36C_1\alpha^3\ln\left(1+\frac{1}{\alpha r}\right)+\frac{6\alpha C_1(-1+6\alpha^2 r^2+3\alpha r)}{r^2(1+\alpha r)}+\frac{1+2\alpha r}{r^2},
\end{eqnarray}
with the Ricci scalar
\begin{eqnarray}
\label{Rr}
R=-12C_2-72C_1\alpha^3\ln\left(1+\frac{1}{\alpha r}\right)+\frac{6\alpha C_1(-1+6\alpha^2 r^2+6\alpha r)(1+2\alpha r)}{r^2(1+\alpha r)^2}+\frac{1}{r^2}.
\end{eqnarray}
From Equation (\ref{sdf}), the entropy of the black hole reads
\begin{eqnarray}
\label{S4}
S=\pi r_{+}^{2}(1+\alpha r_+).
\end{eqnarray}

The energy of the black hole is obtained from Equation (\ref{edf}) as
\begin{eqnarray}
\label{E4}
E&=&\frac{r_+}{2}+\frac{r_+^2}{8}(-6C_1\alpha^2-36\alpha^3C_1r_+ +4C_2r_++3\alpha+6\alpha C_2r_+^2)+\frac{3}{2}\alpha^3C_1r_+^3(2+3\alpha r_+)\ln\left(1+\frac{1}{\alpha r_+}\right)\\\nonumber
&=&C_1-\frac{1}{6\alpha},
\end{eqnarray}
where we used $N(r_+)=0$. For $C_1=0$, Equation (\ref{E4}) reduces to the result in \cite{Zheng:2018fyn}. To guarantee the nonnegativity of the entropy and the energy, we must have new constraints on the parameters: $\alpha\geq -1/r_+$ and $C_1\geq 1/6\alpha$.

\subsubsection{$F=\alpha r^a$}
We now consider a power-law form for $F(r)$, i.e., $F=\alpha r^a$, with constants $a$ and $\alpha$. In this case, the $W(r)$ and $N(r)$ in (\ref{metric3}) were found to be \cite{Amirabi:2015aya}
\begin{eqnarray}
\label{CW}
W=r^{\frac{2a(a-1)}{a+D-2}}N,
\end{eqnarray}
and
\begin{eqnarray}
\label{CN}
N=C_1r^{-\frac{2a^2-6a+6+(2a-5)D+D^2}{a+D-2}}+C_2r^{\frac{2(D-2+2a-a^2)}{a+D-2}}+\frac{(D-3)(a+D-2)^2}{[2a^2-6a+6+(2a-5)D+D^2](D-2+2a-a^2)},
\end{eqnarray}
where $C_1$ and $C_2$ are the integration constants. It returns to the Schwarzschild-de/anti de Sitter solutions for $a=0$ and $\alpha=1$. Function $f(R(r))$ and the Ricci scalar $R$ take the forms, respectively
\begin{eqnarray}
\label{Cf}
f=\frac{2\alpha C_2(D-1)(a-1)(D-2+2a)r^{\frac{a(D-a)}{a+D-2}}}{a+D-2}+\frac{2a\alpha(D-1)(D-3)r^{a-2}}{D-2+2a-a^2},
\end{eqnarray}
\begin{eqnarray}
\label{CR}
 R=-\frac{C_2(D-1)(D-a)(D-2+2a)}{(a+D-2)r^{\frac{2a(a-1)}{a+D-2}}}+\frac{a(D-1)(D-3)(a-2)}{(D-2+2a-a^2)r^2}.
\end{eqnarray}

Note that although $\alpha$ and $a$ are two arbitrary constants, $a$ must satisfy $a\ne2-D,1\pm\sqrt{D-1}$. From Equation (\ref{sdf}) the entropy for this type black hole is
\begin{eqnarray}
\label{S41}
S=\frac{\alpha\pi^{\frac{D-1}{2}}}{2\Gamma(\frac{D-1}{2})} r_{+}^{a+D-2}.
\end{eqnarray}

The energy of the black hole is obtained from Equation (\ref{edf}) as
\begin{eqnarray}
\label{E41}
E=\frac{\pi^\frac{D-3}{2}}{4\Gamma(\frac{D-1}{2})}\dfrac{a_1 r_+^{\frac{D^2-2a-3D+2aD+2}{a+D-2}}+a_2r_+^{\frac{2a^2+2aD+D^2-6a-5D+6}{a+D-2}}}{2(2-2a+a^2-D)[6+2a^2+2a(D-3)-5D+D^2]},
\end{eqnarray}
where
\begin{eqnarray}
 a_1=\alpha C_2(a+D-2)(2-2a+a^2-D)[6+2a^2+2a(D-3)-5D+D^2],
\end{eqnarray}
and
\begin{eqnarray}
 a_2=-\alpha (a+D-2)(D-3)[(3-2D)a^2+(6D-8)a+(D-2)^2].
\end{eqnarray}
{If taking $a=0$ and $\alpha=1$, it is back to Einstein's gravity and Equation (\ref{S41}) reduces to Equation (\ref{ssd}); when $D=3$, $C_2=-\Lambda$, and $C_1=-M$, Equation (\ref{E41}) reduces to Equation (\ref{esd1});
when $D\geq 4$, $C_2=\frac{-2\Lambda}{(D-1)(D-2)}$, and $C_1=\frac{2M}{3-D}$, Equation (\ref{E41}) returns to Equation (\ref{esd2}).} For $D=3$, the solution becomes rather specific since the last term in (\ref{CN}) vanishes for all values of $a$. The function $f(R)$ reads \cite{Amirabi:2015aya}
\begin{eqnarray}
f(R)=a_3R^{\frac{3-a}{2(1-a)}},
\end{eqnarray}
with $a_3=4\alpha C_2(2a^2-a-1)[2C_2(2a+1)(a-3)]^{\frac{a-3}{2(1-a)}}(a+1)^{\frac{1+a}{2(1-a)}}$ and $a\ne 0$. $f(R)$ is a constant for $a=3$ and it is un-physical for $a=1$. The entropy (\ref{S41}) and the energy (\ref{E41}) respectively reduce to
\begin{eqnarray}
S=\frac{\alpha}{2}\pi r_+^{a+1}=\frac{\alpha}{2}\pi\left(-\frac{C_1}{C_2}\right)^{\frac{(a+1)^2}{4a+2}},
\end{eqnarray}
and
\begin{eqnarray}
E=\frac{\alpha C_2(a+1)}{8}r_+^{\frac{4a+2}{a+1}}=-\frac{\alpha C_1(a+1)}{8},
\end{eqnarray}
with $r_+=(\frac{-C_1}{C_2})^{\frac{a+1}{4a+2}}$. The nonnegativity of the entropy gives constraints on the parameters: $C_1(a+1)\leq 0$, and $S\geq 0$ gives $\alpha\geq 0$. For $a=1/3$, we have $f\sim R^2$, $S=\frac{\alpha}{2}\pi r_+^{4/3}=\frac{\alpha}{2}\pi\left(-\frac{C_1}{C_2}\right)^{\frac{8}{15}}$ and $E=-\frac{\alpha C_1}{6}$.

For $D\geq 4$ and $C_2=0$, the function $f(R)$ takes the form \cite{Amirabi:2015aya}
\begin{eqnarray}
f(R)=a_4R^{1-\frac{a}{2}},
\end{eqnarray}
where {\bf $a_4=2\alpha(a-2)^{\frac{a}{2}-1}\left[\frac{a(D-1)(D-3)}{D-2+2a-a^2}\right]^{\frac{a}{2}}$.} The entropy (\ref{S41}) and the energy (\ref{E41}) respectively reads
\begin{eqnarray}
\label{S44}
S&=&\frac{\alpha\pi^{\frac{D-1}{2}}}{2\Gamma(\frac{D-1}{2})} r_{+}^{a+D-2}\nonumber\\
&=&\frac{\alpha\pi^{\frac{D-1}{2}}}{2\Gamma(\frac{D-1}{2})}\left[-\frac{C_1(2a^2-6a+6 + (2a-5)D + D^2)(D-a^2+2a-2)}{(D-3)(D-2+a)^2} \right]  ^{\frac{(D-2+a)^2}{2a^2-6a+6 + (2a-5)D + D^2}},
\end{eqnarray}
\begin{eqnarray}
\label{E44}
E&=&\frac{\pi^\frac{D-3}{2}}{4\Gamma(\frac{D-1}{2})}\dfrac{a_2 r_+^{\frac{2a^2+2aD+D^2-6a-5D+6}{a+D-2}}}{2(2-2a+a^2-D)[6+2a^2+2a(D-3)-5D+D^2]}\nonumber\\
&=& \frac{\pi^\frac{D-3}{2}\alpha C_1}{8\Gamma(\frac{D-1}{2})}\frac{(3-2D)a^2+(6D-8)a+(D-2)^2}{2-a-D},
\end{eqnarray}
with $r_+=\left[-\frac{(D-3)(D-2+a)^2}{C_1[2a^2-6a+6 + (2a-5)D + D^2](D-a^2+2a-2)} \right] ^{-\frac{D-2+a}{2a^2-6a+6 + (2a-5)D + D^2}}$. For the case of $\alpha=1$ and $a=0$ the theory returns to $D$-dimensional Einstein's gravity: $r_+=(-C_1)^{\frac{1}{D-3}}$, $S=\frac{\pi^{\frac{D-1}{2}}}{2\Gamma(\frac{D-1}{2})} (-C_1)^{\frac{D-2}{D-3}}$, and  $E=\frac{\pi^\frac{D-3}{2}}{8\Gamma(\frac{D-1}{2})}(D-2) r_+^{D-3}=-\frac{\pi^\frac{D-3}{2}(D-2)C_1}{8\Gamma(\frac{D-1}{2})}$. For $a=-2$, we get $f\sim R^2$, $S=\frac{\alpha\pi^{\frac{D-1}{2}}}{2\Gamma(\frac{D-1}{2})}\left[-\frac{C_1(D^2-9D+26)(D-10)}{(D-3)(D-4)^2} \right]^{\frac{(D-4)^2}{D^2-9D+26}}$, and $E=\frac{\pi^\frac{D-3}{2}\alpha C_1}{8\Gamma(\frac{D-1}{2})}\frac{D^2-24D+32}{4-D}$, obviously $D\neq 4, 10$. the nonnegativity of the entropy and the energy give new constraints on the parameters: $\alpha\geq 0$ and $C_1(D^2-24D+32)\leq 0$.

\section{conclusions and discussions}
We have discussed whether the new horizon-first law still holds in higher-dimensional $f(R)$ gravity. We have derived the general formulas to calculate the entropy and the energy of a general spherically-symmetric and static $D$-dimensional black hole in $f(R)$ theories, which can be obtained by using other methods. It gives a new method to rapidly compute the entropy and the energy of the black hole in $f(R)$ theory. For applications, we have calculated the entropy and the energy of some black holes with constant Ricci curvature or with non-constant Ricci curvature in some interesting $f(R)$ theory by using these formulas, the nonnegativity of the entropy and the energy give new constraints on the parameters. Except for the case discussed in \cite{Nashed} where $F(R)=0$, it is valuable to apply this procedure to other modified gravitational theories.
\begin{acknowledgments}
This study is supported in part by Hebei Provincial Natural Science Foundation of China (Grant No. A2014201068).
\end{acknowledgments}

\bibliographystyle{ieeetr}
\bibliography{ref}

\end{document}